\documentclass[%
 reprint,
superscriptaddress,
 amsmath,amssymb,
 aps,
prb,
]{revtex4-2}

\usepackage{xcolor}
\usepackage{graphicx}% Include figure files
\usepackage{dcolumn}% Align table columns on decimal point
\usepackage{bm}% bold math
\usepackage{caption}
\usepackage{subcaption}
\usepackage{todonotes}
\usepackage{soul}
\usepackage{comment}

\newcommand{\uu}[1]{\ensuremath{\,\mathrm{#1}}}
\usepackage{gensymb} % \degree
\usepackage{hyperref}
\hypersetup{
    colorlinks=true,
    linkcolor=green,
    urlcolor=cyan,
}

\begin{document}

%%%%%%%%%%%%%%%%%%%%%%%%%%%%%%%%%%%%%%%%%%%%%%%%%%%%%%%%%%%%%%%%%%%%%%%%%%%%%%%%%%%%%

% \title{Hydrogen adsorption on single atom doped grapgene and a 8-0 carbon nanotube}
\title{Revealing trends in catalytic activity of adatoms for hydrogen adsorption on carbon: A case study of graphene and carbon nanotube}

%%%%%%%%%%%%%%%%%%%%%%%%%%%%%%%%%%%%%%%%%%%%%%%%%%%%%%%%%%%%%%%%%%%%%%%%%%%%%%%%%%%%%

\author{Thomas Leiner}
\email{thomas.leiner@unileoben.ac.at}
\affiliation{Department of Materials Science, Montanuniversität Leoben, Leoben, Austria}
\author{David Holec}
\affiliation{Department of Materials Science, Montanuniversität Leoben, Leoben, Austria}
\date{\today}

%%%%%%%%%%%%%%%%%%%%%%%%%%%%%%%%%%%%%%%%%%%%%%%%%%%%%%%%%%%%%%%%%%%%%%%%%%%%%%%%%%%

\begin{abstract}
The increasing demand for sustainable energy solutions necessitates advancements in hydrogen storage technologies. This study investigates the hydrogen adsorption characteristics of graphene and an (8,0) carbon nanotube (CNT) decorated with adatoms of various elements. Using molecular dynamics (MD) simulations and the universal interatomic potential 'PreFerred Potential' (PFP) implemented in the Matlantis framework, we explore the hydrogen storage capabilities of these doped carbon structures at 77\,K. We analyze the adsorption efficiency based on the position of adatoms (top, bridge, and hollow sites) and find that the group II elements, such as calcium and strontium, exhibit significant hydrogen uptake. Additionally, light elements like lithium and sodium demonstrate enhanced gravimetric hydrogen storage due to their low atomic mass. Our findings provide insights into the potential of doped graphene and CNTs for efficient hydrogen storage applications.

\end{abstract}

%%%%%%%%%%%%%%%%%%%%%%%%%%%%%%%%%%%%%%%%%%%%%%%%%%%%%%%%%%%%%%%%%%%%%%%%%%%%%%%%%%%


\maketitle
%%%%%%%%%%%%%%%%%%%%%%%%%%%%%%%%%%%%%%%%%%%%%%%%%%%%%%%%%%%%%%%%%%%%%%%%%%%%%%%%%%%
\section{Introduction}
Due to the ecological problems related to the use of fossil fuels, alternative sources of energy and their storage have to be developed. 
One promising branch of ongoing research focuses on hydrogen. 
The most obvious advantage of using hydrogen instead of carbon-based (fossil) fuels is that after harvesting energy through reacting with oxygen, the ``waste product'' is climate-neutral water instead of carbon dioxide.
Nonetheless, hydrogen technology still has to deal with a number of problems, ranging from economically and ecologically sensible hydrogen production over hydrogen storage in a reasonably small volume and with a high enough ratio of stored hydrogen per storage system mass, as well as developed transport infrastructure to get the hydrogen from the place of its production to the end user. 

This work focuses on revealing trends related to the enhancement of graphene and carbon nanotubes' hydrogen storage capabilities by doping.
Our particular interest is to scan over as many elements of the periodic system as possible. 
While it generally would be possible to simulate these systems with density functional theory (DFT) based approaches, as has been done in previous works~\cite{Holec2018-ye, zhuo2020theoretical, lyu2020overview}, systematically iterating through the entire periodic table for dopants placed in various sites has not been done yet as it would require significant computational resources. 
On the other hand, molecular statics or dynamics would, in principle, also allow for the estimation of such energy-based trends, provided that an interatomic potential exists.
This was not the case until recently, when a paradigm shift to data-driven materials science and machine learning opened the possibility of creating so-called universal potentials~\cite{Riebesell2023-rv}.
In the present work, we employ the 'PreFerred Potential' (PFP) universal potential~\cite{takamoto2022towards, takamoto2023towards}.
Not only does it allow to perform the desired simulations with acceptable computational resources, but it also reveals systematic trends across the periodic table and enables the use of larger (than in DFT) simulation models and including finite temperature effects.

While there exist many hydrogen adsorption studies for doped graphene~\cite{miwa2008hydrogen, tabtimsai2017hydrogen, Holec2018-ye, jain2020functionalized} and 8-0 CNTs~\cite{NAKANO2006125, nagare2012hydrogen, verdinelli2014theoretical, lyu2020overview}, typically they focus on only one or a few dopants. 
Moreover, the conclusions are drawn using different system setups and methodologies, which makes comparing these studies difficult. 
This work aims to directly compare the effect of the doping elements across (almost) the entire periodic table with each other.

%%%%%%%%%%%%%%%%%%%%%%%%%%%%%%%%%%%%%%%%%%%%%%%%%%%%%%%%%%%%%%%%%%%%%%%%%%%%%%%%%%%
\section{Methods}

%%%%%%%%%%%%%%%%%%%%%%%%%%%%%%%%%%%%%%%%%%%%%%%%%%%%%%%%%%%%%%%%%%%%%%%%%%%%%%%%%%%
\subsection{Structural models}

\begin{figure}[b]
    \centering
    \begin{minipage}[b]{0.45\linewidth}
        \centering
        \includegraphics[width=\linewidth]{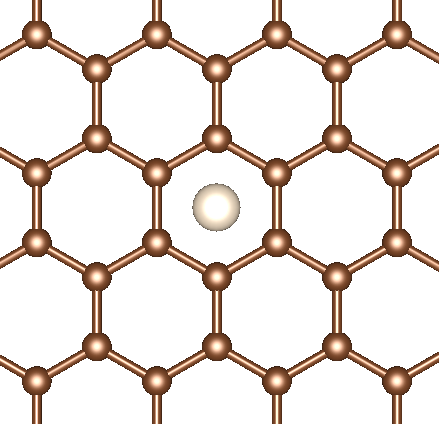}
        \subcaption{Hollow site}
    \end{minipage}
    \hspace{0.05\linewidth}
    \begin{minipage}[b]{0.45\linewidth}
        \centering
        \raisebox{1.5ex}{\includegraphics[width=\linewidth]{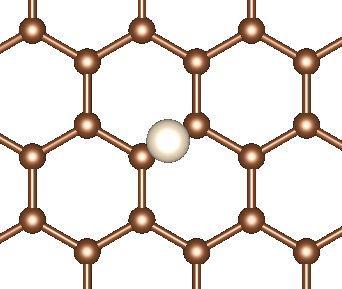}}
        \subcaption{Bridge site}
    \end{minipage}
    \vspace{0.3cm} % Add vertical space between rows
    \begin{minipage}[b]{0.45\linewidth}
        \centering
        \includegraphics[width=\linewidth]{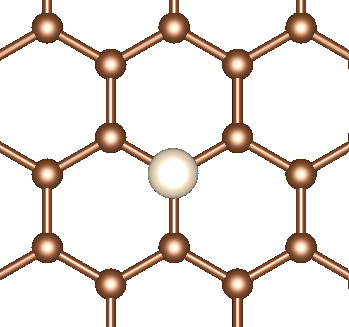}
        \subcaption{Top site}
    \end{minipage}
    \hspace{0.05\linewidth}
    % \begin{minipage}[b]{0.45\linewidth}
    %     \centering
    %     \raisebox{4.5ex}{\includegraphics[width=\linewidth]{figs/graphene_substitutional.png}}
    %     \subcaption{Substitutional position}
    % \end{minipage}
    \caption{Structures with graphene basis: (a)~hollow site, (b)~bridge site, (c)~top site.%, (d)~substitutional position.
    The adatom is represented by the white atom.
    }
    \label{fig:graphene_structures}
\end{figure}

As the systems of interest, we chose two fundamental carbon structures, a graphene sheet and a carbon nanotube (CNT) rolled in the 8-0 configuration~\cite{iijima1991helical}.

On a graphene sheet, it is possible to place an additional adatom in three different (high-symmetry) places: in the middle of the carbon hexagons (hollow site), above a carbon-carbon bond (bridge site), or on top of a carbon atom (top site), as visualized in Fig.~\ref{fig:graphene_structures}a--c. 
The simulation model of the graphene sheet was contained in an $86\times75\times26.\,\text{\AA}^3$ simulation cell, corresponding to a $35\times35$ supercell of graphene (2450 C atoms).
In total, 49 doping atoms were placed in the desired positions (hole, bridge, top), followed by an introduction of 392 hydrogen molecules placed on a regular grid in a vacuum region away from the graphene sheet.
The resulting model contained 3283 atoms.% (3234 in the case of substitutional doping).

\begin{figure}[tbp]
    \centering
    \begin{minipage}[t]{0.45\linewidth}
        \centering
        \includegraphics[width=\linewidth]{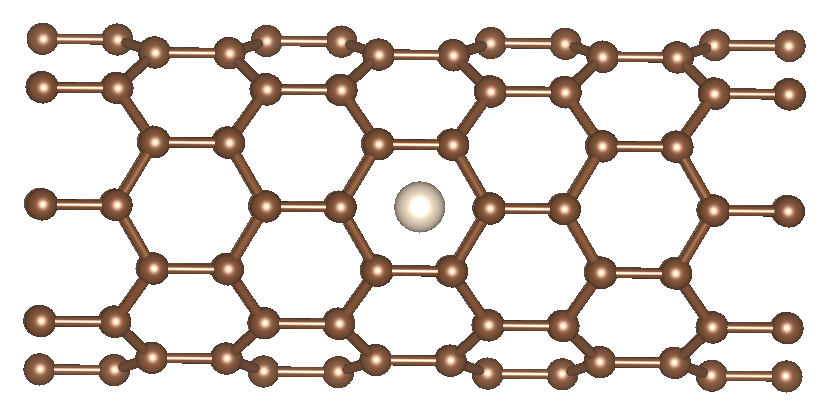}
        \subcaption{Hole position}
    \end{minipage}
    \hspace{0.05\linewidth}
    \begin{minipage}[t]{0.45\linewidth}
        \centering
        \includegraphics[width=\linewidth]{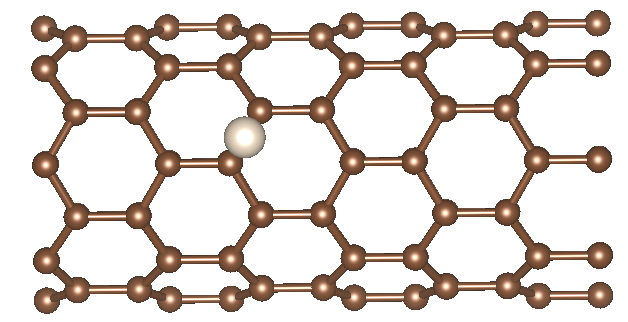}
        \subcaption{Bridge position diagonal}
    \end{minipage}
    \vspace{0.05\linewidth}
    \begin{minipage}[t]{0.45\linewidth}
        \centering
        \includegraphics[width=\linewidth]{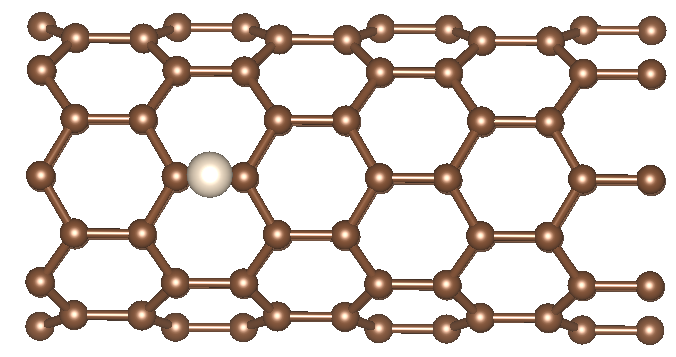}
        \subcaption{Bridge position horizontal}
    \end{minipage}
    \hspace{0.05\linewidth}
    \begin{minipage}[t]{0.45\linewidth}
        \centering
        \includegraphics[width=\linewidth]{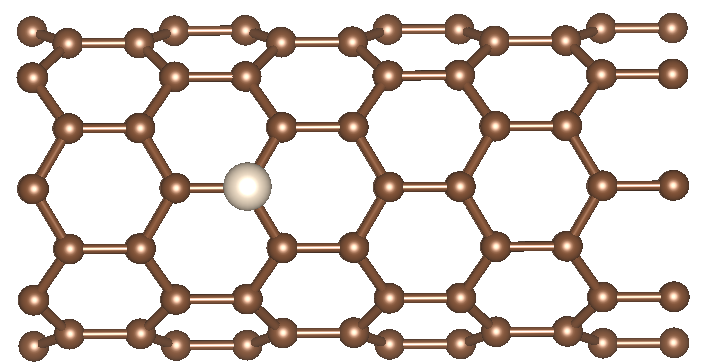}
        \subcaption{Top position}
    \end{minipage}
    \vspace{0.05\linewidth}
    % \begin{minipage}[t]{0.45\linewidth}
    %     \centering
    %     \includegraphics[width=\linewidth]{figs/8-0CNP_with_element_substitutional.png}
    %     \subcaption{Substitutional position}
    % \end{minipage}
    \caption{Structures with CNT basis: (a)~hole position, (b)~bridge position diagonal, (c)~bridge position horizontal, (d)~top position.%, (e)~substitutional position.
    The dopant is represented by the white atom.    
    }
    \label{fig:CNT_structures}
\end{figure}

Analogous sites were also chosen for the placement of the adatoms in the CNT case (Fig.~\ref{fig:CNT_structures}), with the exception that there are two unequal carbon-carbon bonds and therefore two different bridge sites: one which we designate as diagonal (Fig.~\ref{fig:CNT_structures}b) and the other horizontal (Fig.~\ref{fig:CNT_structures}c).

The 8-0 CNT with a radius of $R=3.15\,\text{\AA}$ and a C-C bond lenght of $1.421\,\text{\AA}$ was created using an online tool~\cite{tubegen34}.
Four identical axes-aligned CNTs, each composed of 196 atoms, were placed in a $40\times40\times26.6\,\text{\AA}^3$ simulation box. 
%\todo[inline]{Did you use ASE or any other package for this? C--C bond length?TL: I used a CNT generator (https://turin.nss.udel.edu/research/tubegenonline.html) with C-C bond lenght of 1.4210,Will add the citation.} 
Subsequently, 16 adatoms were introduced (4 on each CNT), resulting in a similar atomic concentration (2\,at.\%) as in the graphene case.
Finally, the empty space was filled with 128 H$_2$ molecules randomly placed,
thereby yielding models with 1040 atoms.% (1024 in the case of substitutional placement)

\begin{figure}[tbp]
    \centering
    \begin{minipage}[t]{0.46\linewidth}
        \centering
        \includegraphics[width=\linewidth]{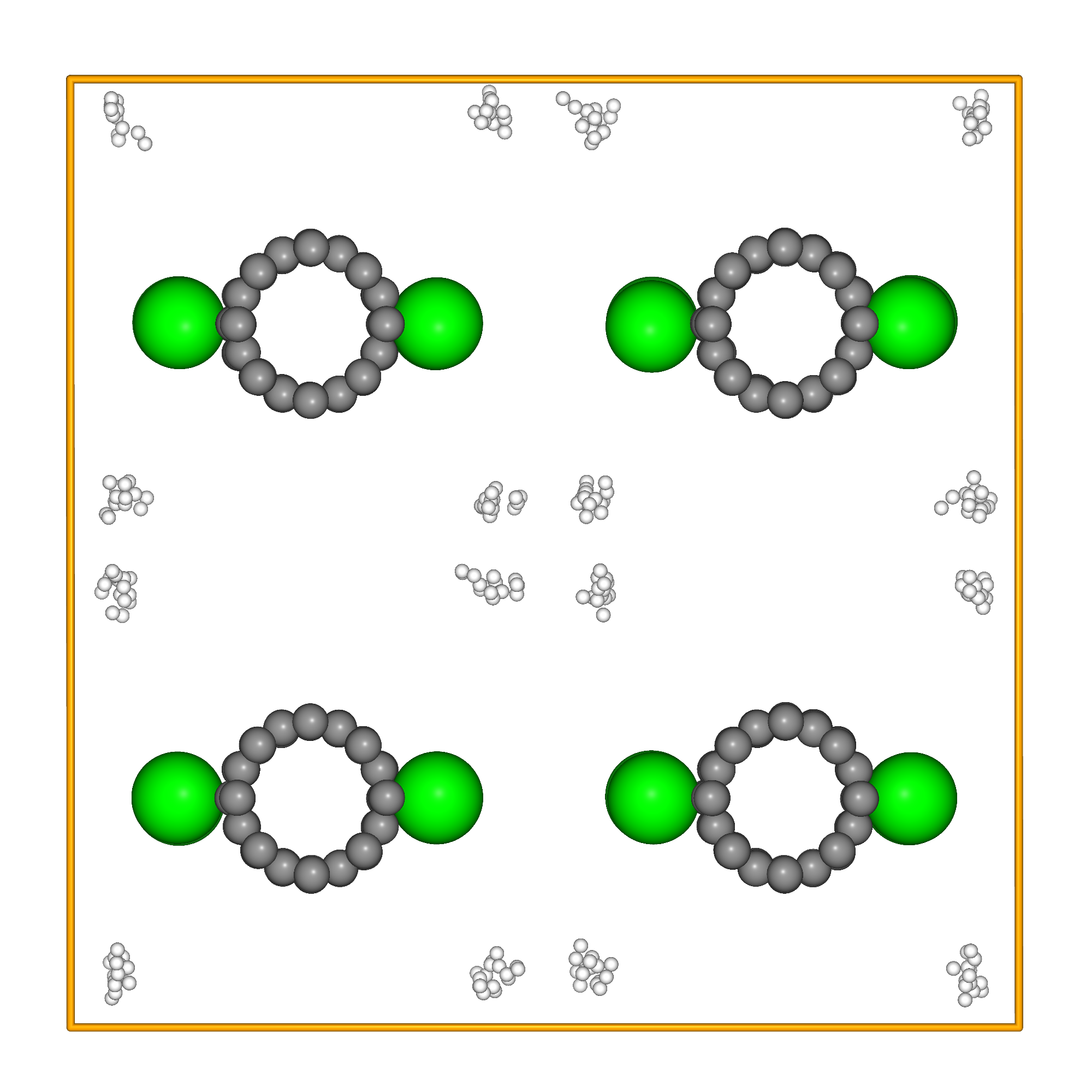}
        \subcaption{Start of the simulation}
    \end{minipage}
    \hspace{0.05\linewidth}
    \begin{minipage}[t]{0.45\linewidth}
        \centering
        \includegraphics[width=\linewidth]{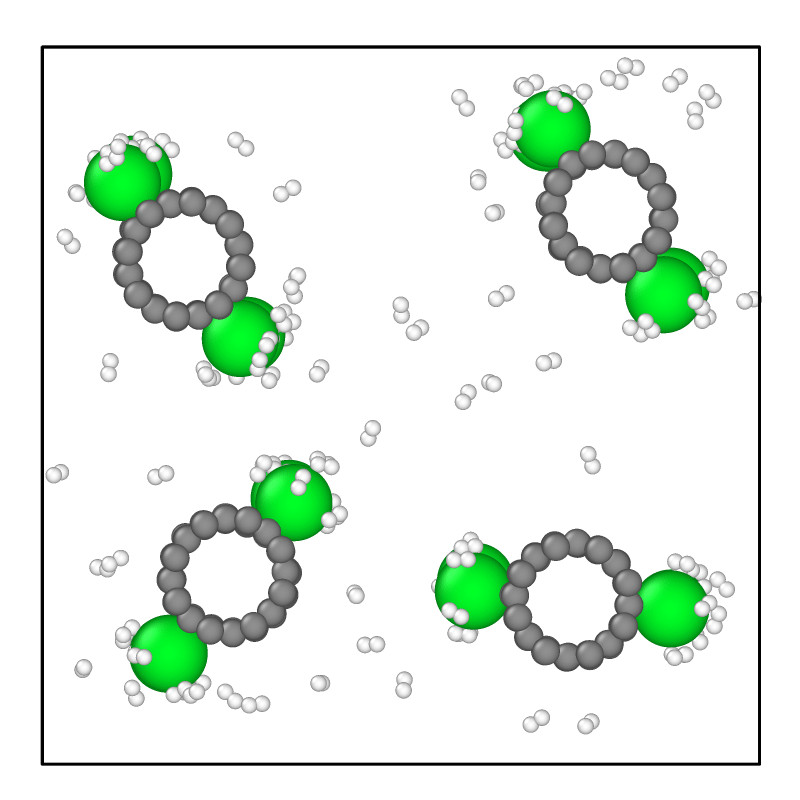}
        \subcaption{End of the simulation}
    \end{minipage}
    \caption{Simulation cell with four 8-0 CNTs(grey) with adsorbed Strontium atoms(green) in the hole position. Adsorbed and free hydrogen molecules are shown in white.}
\label{fig:Sr_example}
\end{figure}

In all cases, the adatoms were evenly spaced to minimize the influence that a dopant-dopant interaction could have. 
An example of a simulation cell is given in Fig.~\ref{fig:Sr_example}.
Periodic boundary conditions were applied in all three spatial directions.

%This setup allowed the adsorbed metal atoms to stay in the positions they were placed without desorption from the carbon surface \hl{or surface migration -> migrate}.

%\todo{Rework the moving of atoms part }
%%%%% YOU DON'T DISCUSS THOSE RESULTS SO I PROPOSE TO REMOVE THEM COMPLETY. TL: I agree
% The substitutional placements of the dopants lead in the most cases to the atoms adsorbing above the graphene sheet in what is now a vacancy in the graphene sheet. As this would better fit to work of defected carbon structures, it will be part of future work.

%\begin{figure}[htbp]
%    \centering
%    \includegraphics[width=0.5\linewidth]{figs/8-0CNT_adsorption_on_Sr_in_hole.jpg}
%    \caption{Simulation cell with four 8-0 CNTs(grey) with adsorbed Strontium atoms(green) in the hole position. Adsorbed and free hydrogen molecules in white.}
%\label{fig:Sr_example}
%\end{figure}

%%%%%%%%%%%%%%%%%%%%%%%%%%%%%%%%%%%%%%%%%%%%%%%%%%%%%%%%%%%%%%%%%%%%%%%%%%%%%%%%%%%
\subsection{Simulation environment}
For our work, we chose a universal potential called `Preferred Potential' (PFP) provided by the Preferred Computational Chemistry (PFCC).
It describes interactions between 72 different elements of the periodic table and is made accessible with a cloud simulation platform called Matlantis\textsuperscript{TM}~\cite{takamoto2022towards, takamoto2023towards}. 

PFP is a neural network potential (NNP) based on the TeaNet architecture~\cite{takamoto2022teanet} with an additional Morse-style two-body potential term for short-range repulsion.
Its training and validation dataset was generated using the Vienna Ab-initio Simulation Package (VASP)~\cite{kresse1996efficiency, kresse1996efficient} with a Perdew–Burke–Ernzerhof (PBE) generalized-gradient approximation for the exchange-correlation functional~\cite{perdew1996generalized}.
The projector-augmented wave (PAW)~\cite{blochl1994projector, kresse1999ultrasoft} method and plane-wave basis set were used together with the kinetic energy cutoff set to 520\,eV. 
Gaussian smearing was applied with a smearing width of 0.05\,eV. 
The $k$-point density was set to 1000 $k$-points per reciprocal atom~\cite{takamoto2022towards}. 
The energy of the system, atomic forces, and atomic charges are used for the training procedure.
For additional information about the generation of the potential, see the original papers~\cite{takamoto2022towards, takamoto2023towards}.
During all PFP calculations, a van-der-Waals D3 correction by Grimme et al.~\cite{Grimme2010-yv} was applied.

Motivated by experimental practice~\cite{KOSTOGLOU2021294}, the temperature of our molecular dynamics simulations was set to 77\,K.
The MD simulations were performed using an $NVT$ ensemble, using a Berendsen thermostat integrator~\cite{berendsen1984molecular}.
They ran for 50,000 timesteps, each 1\,fs long, yielding a total simulation time of 50\,ps.

All evaluations were done using the integrated Jupyter/python-based Matlantis environment, including also the ASE tools~\cite{ASE}.
A hydrogen atom was considered adsorbed to an adatom if their mutual distance was shorter than $3\,\text{\AA}$.

%%%%%%%%%%%%%%%%%%%%%%%%%%%%%%%%%%%%%%%%%%%%%%%%%%%%%%%%%%%%%%%%%%%%%%%%%%%%%%%%%%%
\section{Results and discussion}
%%%%%%%%%%%%%%%%%%%%%%%%%%%%%%%%%%%%%%%%%%%%%%%%%%%%%%%%%%%%%%%%%%%%%%%%%%%%%%%%%%%

In order to have representative results, we made sure that the simulations reached an equilibrated state after the 50,000 timesteps, as measured by a constant amount of hydrogen molecules adsorbed to the dopant elements (i.e., on average, an equal amount of hydrogen adsorbs and desorbs from the adatom sites).

\begin{figure}[ht]
    \centering
    \includegraphics[width=\linewidth]{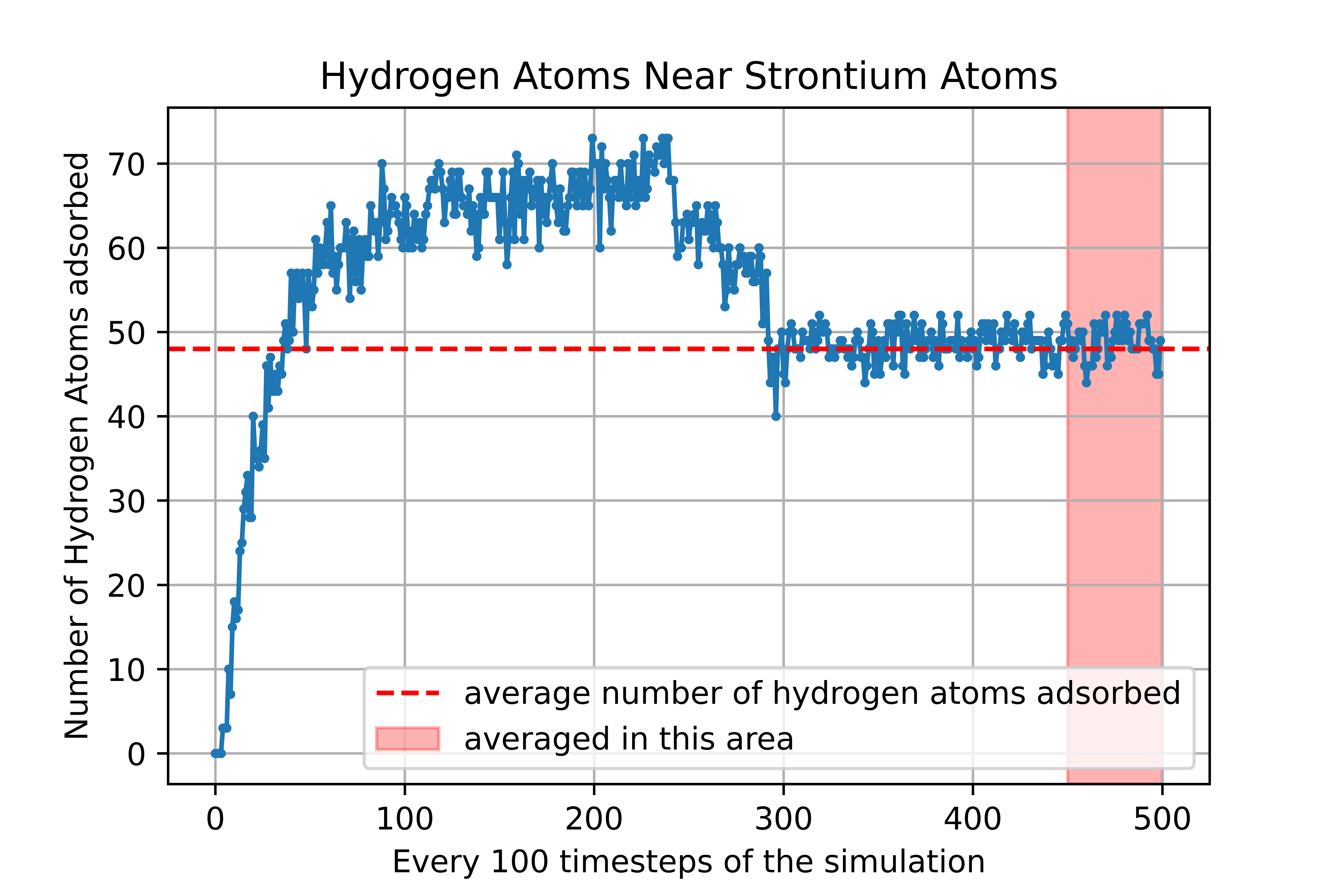}
    \caption{Number of hydrogen atoms adsorbed to Strontium over the entirety of the simulation, with a datapoint every hundred timesteps.}
\label{fig:adsorbed_number}
\end{figure}

To quantify the number of adsorbed hydrogen molecules, we summed up the number of hydrogen atoms within $3\,\text{\AA}$ of the dopant.
The representative number used to extract trends over the period table is then obtained as the time average over the last 5,000 timesteps (5\,ps) of the simulation. 
An example that shows that the amount of adsorbed hydrogen atoms is in thermodynamic equilibrium is seen in Fig. \ref{fig:adsorbed_number}, with the average number of adsorbed hydrogen atoms on Strontium adatoms being $\approx48$ (red dashed line) with fluctuations of $\approx 5\%$ at the end of the simulation (red shaded area).

%Figure~\ref{fig:adsorption_heatmap} shows the hydrogen adsorption depending on the dopant element on the graphene sheet on a color scheme, where white means zero adsorption observed, while dark blue shows the most hydrogen adsorbed. Depending on the placement of the dopant atom, various amounts of hydrogen get adsorbed to the dopant atom, with most adsorption observed for group two elements, Ca and Sr, and various transition metals. Fig.~\ref{fig:adsorption_heatmap_cnt} shows the analogue of Fig.~\ref{fig:adsorption_heatmap} for the simulated 8-0 CNT. ---> replace this paragrap wit te description of the new average heatmap

\begin{figure}
    \centering
    \includegraphics[width=1\linewidth]{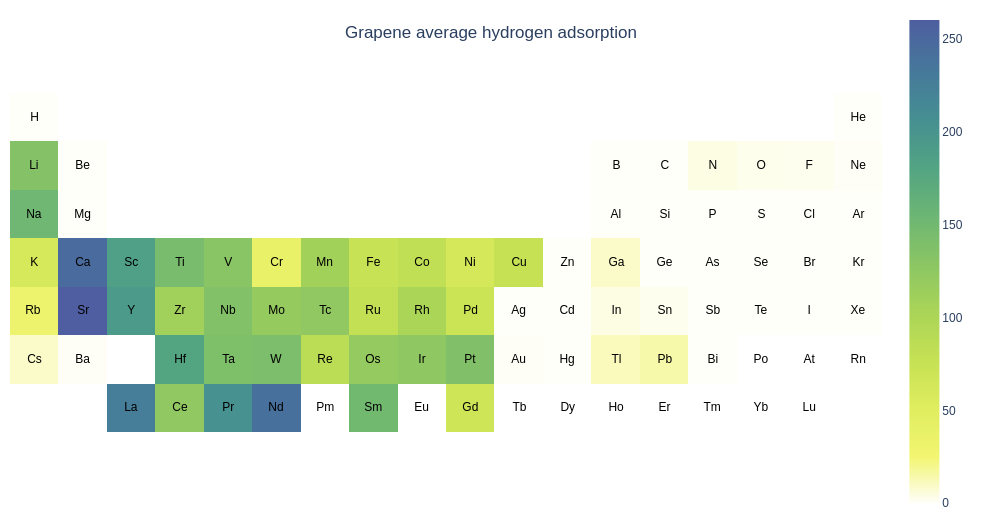}
    \caption{Adsorption heatmap on graphene sheet with active adatoms.}
    \label{fig:graohene_av_ads}
\end{figure}

\begin{figure}
    \centering
    \includegraphics[width=1\linewidth]{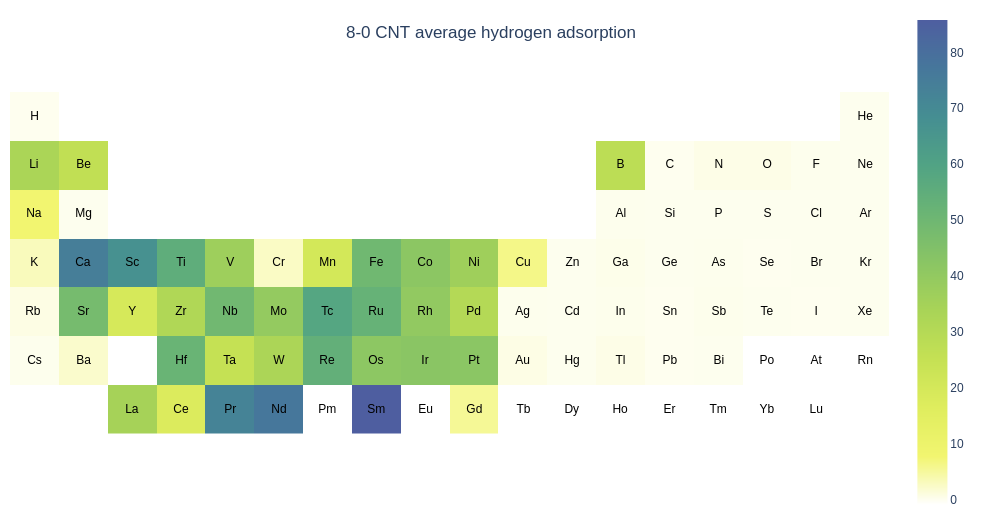}
    \caption{Adsorption heatmap on a 8-0 CNT with active adatoms.}
    \label{fig:8-0_av_ads}
\end{figure}

The case-resolved average adsorption capacities of individual adatoms are given in Supplementary Material.
Since the different sites for graphene and for the CNT show qualitatively the same trends, we averaged them and presented them in the form of heatmaps in Figs.~\ref{fig:graohene_av_ads} and \ref{fig:8-0_av_ads}. 
The best performing adatoms, i.e. those binding the highest number of H atoms, are alkali earth metals (calcium, strontium), some early transition metals (scandium, titanium, hafnium, (yttrium), (niobium)) and some lanthanides (praseodymium, neodymium, (lanthanum), (samarium)).
Although promising, technetium is rather impractical due to its radioactivity.

\begin{figure}[ht]
    \centering
    \includegraphics[width=\linewidth]{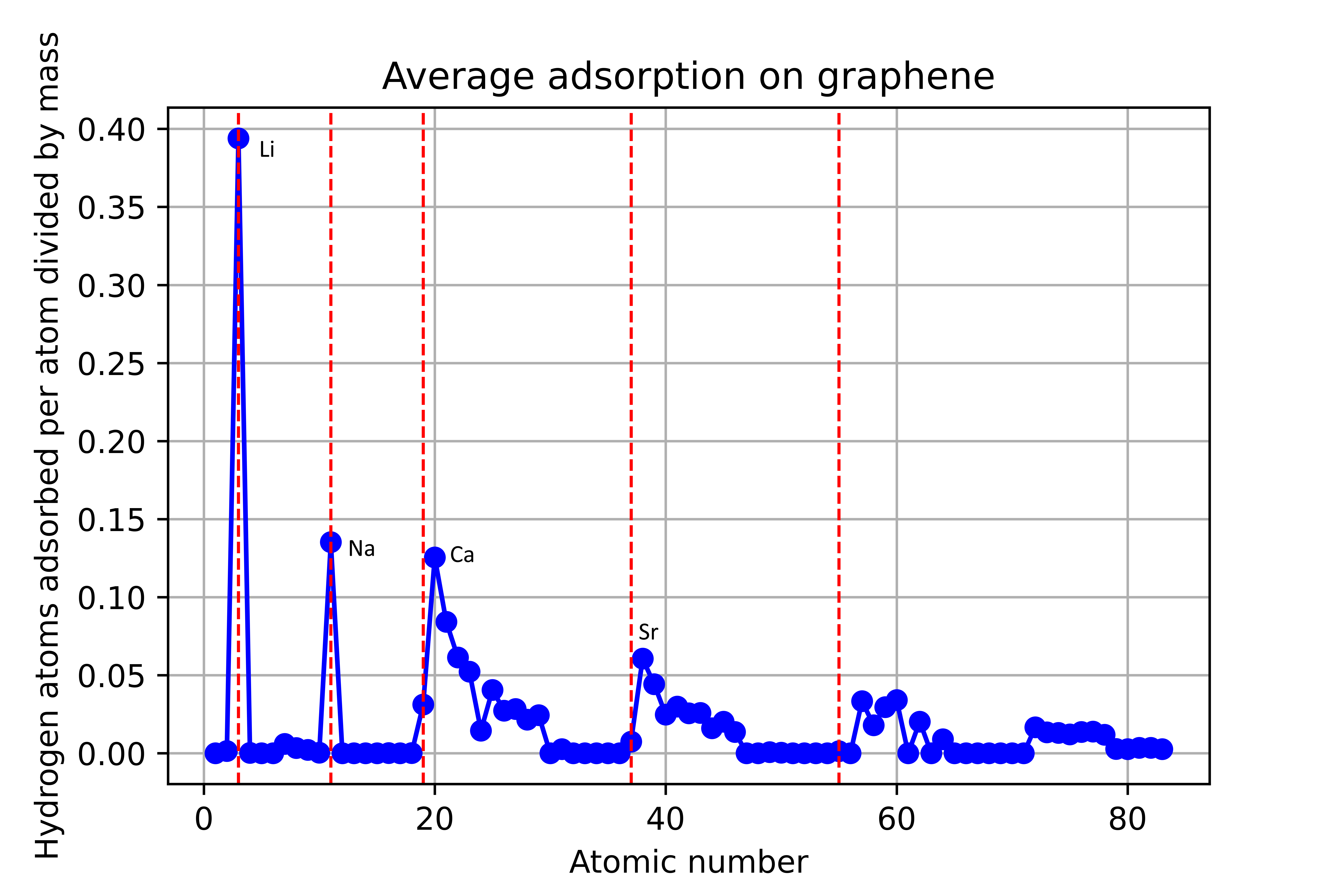}
    \caption{Gravimetric uptake of hydrogen on graphene with active adatoms.}
\label{fig:Mass weighted hydrogen adsorption per element for graphene}
\end{figure}

\begin{figure}[ht]
    \centering
    \includegraphics[width=\linewidth]{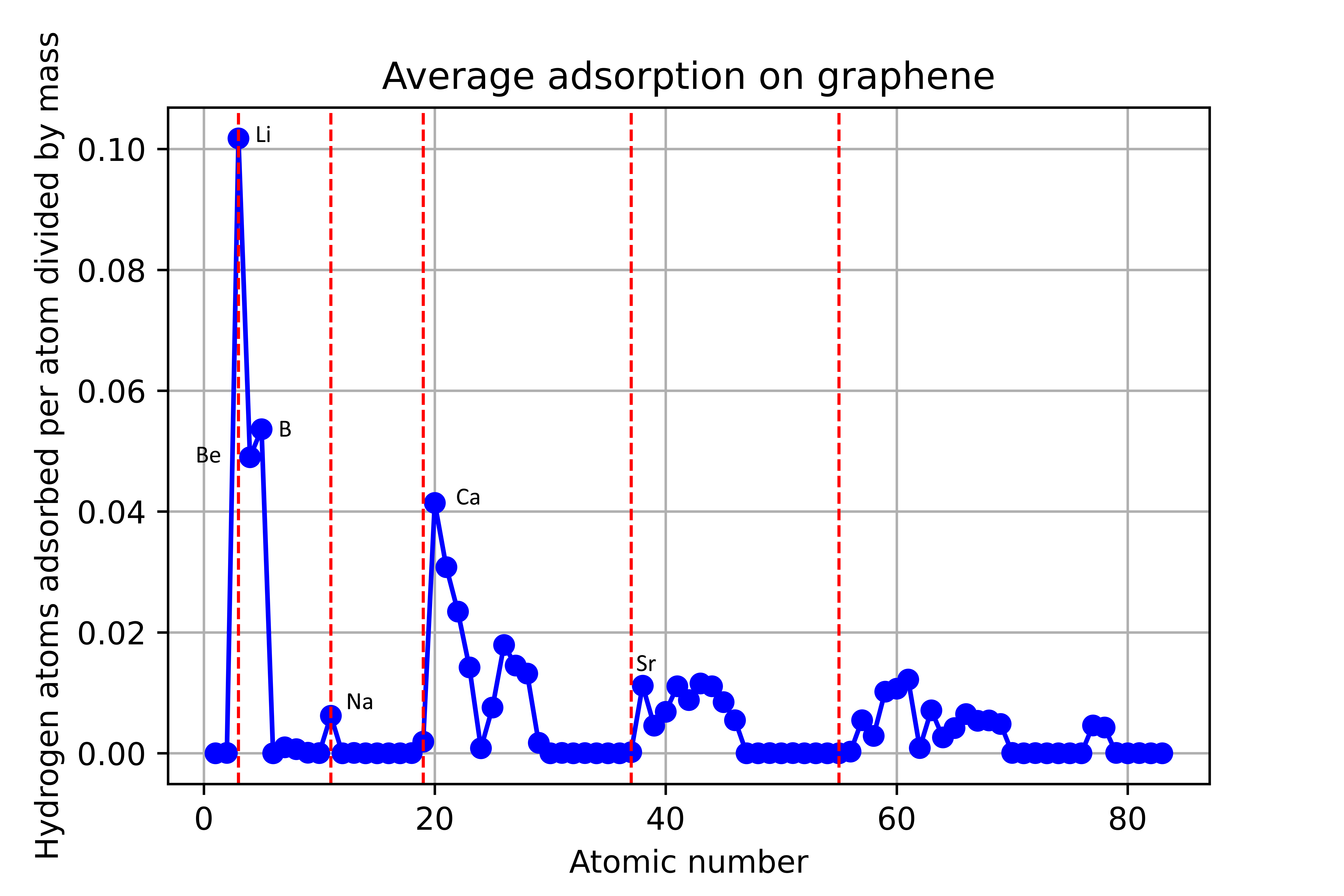}
    \caption{Gravimetric uptake of hydrogen on 8-0 CNT  with active adatoms.}
\label{fig:Mass weighted hydrogen adsorption per element for a 8-0 CNT}
\end{figure}

Figures~\ref{fig:Mass weighted hydrogen adsorption per element for graphene} and \ref{fig:Mass weighted hydrogen adsorption per element for a 8-0 CNT} show the average gravimetric uptake of hydrogen for the graphene and CNT cases. 
This is an important parameter in hydrogen storage applications, as the system size and weight are critical for practical hydrogen storage applications.
To calculate it, the number of absorbed H atoms was divided by the atomic mass of the adatom elements. 
This shifts the spotlight towards light elements, as seen by the peaks for lithium, sodium, and calcium, which thereby outperform the strongly attractive but heavy adatoms such as lanthanum, neodymium, or samarium.
Significant differences between graphene and the CNT are the higher peak for sodium in Fig.~\ref{fig:Mass weighted hydrogen adsorption per element for graphene} and the high values for boron and beryllium in Fig.~\ref{fig:Mass weighted hydrogen adsorption per element for a 8-0 CNT}. 
Last but not least, strontium seems to be significantly enhancing the H adsorption only in the case of graphene (Fig.~\ref{fig:Mass weighted hydrogen adsorption per element for graphene}).
As a general trend for both graphene and CNT cases, a declining trend in gravimetric capacity is predicted within each individual period (individual periods marked with dashed red lines).
Note also the different $y$-axis scales in both figures, suggesting the flat graphene is 3--4 times more efficient than the CNT.

Several related works have been published previously.
A recent study by Myrzakhmetov et al. \cite{MYRZAKHMETOV2025413} shows a physical bonding of up to $8.8\uu{wt.\%}$ hydrogen on Li-doped graphene, when lithium is placed in the hollow site. 
This placement of lithium also shows the highest hydrogen adsorption found in this study (see SI Fig.~1). 
A significant hydrogen adsorption on beryllium-decorated CNTs has been reported by Ghosh et al. \cite{GHOSH201724237}. 
Boron is typically considered as a substitutional element both in graphene \cite{zhou2009adsorption} and CNTs \cite{SAWANT202139297}.
Sawant et al. \cite{SAWANT202139297} reported on enhanced hydrogen storage for boron-decorated a CNT (BCNT). 
Similarly, Lee et al.\cite{lee2009calcium, lee2010calcium} showed hydrogen storage capability enhancement for calcium-doped graphene and CNTs.
All these primarily experimental studies are in line with our predictions on the effect of chemical doping of carbon-based nanostructures.

Two effects are reported in the literature concerning the hydrogen uptake of doped carbon structures, namely hydrogen spillover~\cite{pyle2016hydrogen} and the Kubas-type interaction~\cite{kubas1988molecular}. 

Brzhezinskaya et al. \cite{brzhezinskaya2014electronic, brzhezinskaya2019new} report chemically bound hydrogen on the CNTs after hydrogenation at $p\approx5\uu{GPa}$ and $T\approx500\uu{^{\circ}C}$ and Talyzin et al. \cite{talyzin2011hydrogenation} report a chemical hydrogenation at temperatures at and above $400\uu{^{\circ}C}$ at a pressure of $50\uu{bar}$ ($0.005\uu{GPa}$. 

However, at ambient conditions, the hydrogen spillover is mainly reported for defective carbon structures \cite{pyle2016hydrogen}.
In agreement with this observation, we did not observe it during our simulations employing pristine carbon structures. 
Another reason for not observing the hydrogen spillover might be the limited simulation time, which, together with the limited kinetics at a low temperature of $77\,\text{K}$, does not allow for spillover to occur. 
This study could be expanded for future research by incorporating defective graphene planes, such as those with Stone–Wales defects or sheet edges, to investigate hydrogen spillover more closely.
A study of the electronic structure of a local environment representing an adatom with an adsorbed hydrogen could reveal the role of the Kubas-type interaction in the adsorption enhancement.
For this, an electronic-structure code would be necessary, which goes beyond the capabilities of the used universal potential.
Clustering was reported for transition metal dopants \cite{lambie2022clustering, sun2005clustering}, which poses an additional technical problem in synthesizing the material, when considering the gravimetric and total hydrogen uptake capability of the materials. 
Therefore, this topic would also be interesting to include in the follow-up studies.%%%%%%%%%%%%%%%%%%%%%%%%%%%%%%%%%%%%%%%%%%%%%%%%%%%%%%%%%%%%%%%%%%%%%%%%%%%%%%%%%%%

\section{Conclusions}
Hydrogen technologies are considered to be part of the solution to the problems related to climate change.
In this paper, we present a systematic theoretical investigation of the catalytic activity of adatoms enhanced hydrogen uptake by carbon-based nanostructures, namely graphene and 8-0 CNT.
To do so, we performed a molecular dynamics study employing a recently published neural network interatomic potential fitted to DFT-GGA data.
Our calculations clearly identified multiple atomic species that can lead to improved gravimetric hydrogen uptake. 

The most promising element is lithium for both considered carbon allotropes, followed by calcium and scandium.
Other elements show enhancement only for certain carbon configurations. While sodium is the second-best dopant for graphene, beryllium and boron are predicted to be effective in the case of the 8-0 CNT.

These predictions are in line with previous theoretical and experimental reports on the effect of chemical doping of graphene and CNTs.
Although differences in simulation setups and experimental conditions may hinder obtaining a quantitative resemblance, this work provides a direct comparison of the role of the vast majority of elements in the periodic system, when used as dopants in graphene and the 8-0 CNT.
Therefore, the present results aim to provide useful data for steering the development of doping-activated carbon nanostructures with enhanced hydrogen-storage capacity.

%%%%%%%%%%%%%%%%%%%%%%%%%%%%%%%%%%%%%%%%%%%%%%%%%%%%%%%%%%%%%%%%%%%%%%%%%%%%%%%%%%%
\section*{Acknowledgements}
This work is a part of the Strategic Core Research Area (SCoRe A+ Hydrogen and Carbon) and is receiving financial support from Montanuniversität Leoben, Austria. Data will be made available upon a reasonable request.
\bibliography{refs}

\begin{thebibliography}{37}%
\makeatletter
\providecommand \@ifxundefined [1]{%
 \@ifx{#1\undefined}
}%
\providecommand \@ifnum [1]{%
 \ifnum #1\expandafter \@firstoftwo
 \else \expandafter \@secondoftwo
 \fi
}%
\providecommand \@ifx [1]{%
 \ifx #1\expandafter \@firstoftwo
 \else \expandafter \@secondoftwo
 \fi
}%
\providecommand \natexlab [1]{#1}%
\providecommand \enquote  [1]{``#1''}%
\providecommand \bibnamefont  [1]{#1}%
\providecommand \bibfnamefont [1]{#1}%
\providecommand \citenamefont [1]{#1}%
\providecommand \href@noop [0]{\@secondoftwo}%
\providecommand \href [0]{\begingroup \@sanitize@url \@href}%
\providecommand \@href[1]{\@@startlink{#1}\@@href}%
\providecommand \@@href[1]{\endgroup#1\@@endlink}%
\providecommand \@sanitize@url [0]{\catcode `\\12\catcode `\$12\catcode `\&12\catcode `\#12\catcode `\^12\catcode `\_12\catcode `\%12\relax}%
\providecommand \@@startlink[1]{}%
\providecommand \@@endlink[0]{}%
\providecommand \url  [0]{\begingroup\@sanitize@url \@url }%
\providecommand \@url [1]{\endgroup\@href {#1}{\urlprefix }}%
\providecommand \urlprefix  [0]{URL }%
\providecommand \Eprint [0]{\href }%
\providecommand \doibase [0]{https://doi.org/}%
\providecommand \selectlanguage [0]{\@gobble}%
\providecommand \bibinfo  [0]{\@secondoftwo}%
\providecommand \bibfield  [0]{\@secondoftwo}%
\providecommand \translation [1]{[#1]}%
\providecommand \BibitemOpen [0]{}%
\providecommand \bibitemStop [0]{}%
\providecommand \bibitemNoStop [0]{.\EOS\space}%
\providecommand \EOS [0]{\spacefactor3000\relax}%
\providecommand \BibitemShut  [1]{\csname bibitem#1\endcsname}%
\let\auto@bib@innerbib\@empty
%</preamble>
\bibitem [{\citenamefont {Holec}\ \emph {et~al.}(2018)\citenamefont {Holec}, \citenamefont {Kostoglou}, \citenamefont {Tampaxis}, \citenamefont {Babic}, \citenamefont {Mitterer},\ and\ \citenamefont {Rebholz}}]{Holec2018-ye}%
  \BibitemOpen
  \bibfield  {author} {\bibinfo {author} {\bibfnamefont {D.}~\bibnamefont {Holec}}, \bibinfo {author} {\bibfnamefont {N.}~\bibnamefont {Kostoglou}}, \bibinfo {author} {\bibfnamefont {C.}~\bibnamefont {Tampaxis}}, \bibinfo {author} {\bibfnamefont {B.}~\bibnamefont {Babic}}, \bibinfo {author} {\bibfnamefont {C.}~\bibnamefont {Mitterer}},\ and\ \bibinfo {author} {\bibfnamefont {C.}~\bibnamefont {Rebholz}},\ }\bibfield  {title} {\bibinfo {title} {Theory-guided metal-decoration of nanoporous carbon for hydrogen storage applications},\ }\href {https://doi.org/10.1016/j.surfcoat.2018.07.025} {\bibfield  {journal} {\bibinfo  {journal} {Surf. Coat. Technol.}\ }\textbf {\bibinfo {volume} {351}},\ \bibinfo {pages} {42} (\bibinfo {year} {2018})}\BibitemShut {NoStop}%
\bibitem [{\citenamefont {Zhuo}\ \emph {et~al.}(2020)\citenamefont {Zhuo}, \citenamefont {Zhang}, \citenamefont {Liang}, \citenamefont {Yu}, \citenamefont {Xiao},\ and\ \citenamefont {Li}}]{zhuo2020theoretical}%
  \BibitemOpen
  \bibfield  {author} {\bibinfo {author} {\bibfnamefont {H.-Y.}\ \bibnamefont {Zhuo}}, \bibinfo {author} {\bibfnamefont {X.}~\bibnamefont {Zhang}}, \bibinfo {author} {\bibfnamefont {J.-X.}\ \bibnamefont {Liang}}, \bibinfo {author} {\bibfnamefont {Q.}~\bibnamefont {Yu}}, \bibinfo {author} {\bibfnamefont {H.}~\bibnamefont {Xiao}},\ and\ \bibinfo {author} {\bibfnamefont {J.}~\bibnamefont {Li}},\ }\bibfield  {title} {\bibinfo {title} {Theoretical understandings of graphene-based metal single-atom catalysts: stability and catalytic performance},\ }\href@noop {} {\bibfield  {journal} {\bibinfo  {journal} {Chemical reviews}\ }\textbf {\bibinfo {volume} {120}},\ \bibinfo {pages} {12315} (\bibinfo {year} {2020})}\BibitemShut {NoStop}%
\bibitem [{\citenamefont {Lyu}\ \emph {et~al.}(2020)\citenamefont {Lyu}, \citenamefont {Kudiiarov},\ and\ \citenamefont {Lider}}]{lyu2020overview}%
  \BibitemOpen
  \bibfield  {author} {\bibinfo {author} {\bibfnamefont {J.}~\bibnamefont {Lyu}}, \bibinfo {author} {\bibfnamefont {V.}~\bibnamefont {Kudiiarov}},\ and\ \bibinfo {author} {\bibfnamefont {A.}~\bibnamefont {Lider}},\ }\bibfield  {title} {\bibinfo {title} {An overview of the recent progress in modifications of carbon nanotubes for hydrogen adsorption},\ }\href@noop {} {\bibfield  {journal} {\bibinfo  {journal} {Nanomaterials}\ }\textbf {\bibinfo {volume} {10}},\ \bibinfo {pages} {255} (\bibinfo {year} {2020})}\BibitemShut {NoStop}%
\bibitem [{\citenamefont {Riebesell}\ \emph {et~al.}(2023)\citenamefont {Riebesell}, \citenamefont {Goodall}, \citenamefont {Benner}, \citenamefont {Chiang}, \citenamefont {Deng}, \citenamefont {Lee}, \citenamefont {Jain},\ and\ \citenamefont {Persson}}]{Riebesell2023-rv}%
  \BibitemOpen
  \bibfield  {author} {\bibinfo {author} {\bibfnamefont {J.}~\bibnamefont {Riebesell}}, \bibinfo {author} {\bibfnamefont {R.~E.~A.}\ \bibnamefont {Goodall}}, \bibinfo {author} {\bibfnamefont {P.}~\bibnamefont {Benner}}, \bibinfo {author} {\bibfnamefont {Y.}~\bibnamefont {Chiang}}, \bibinfo {author} {\bibfnamefont {B.}~\bibnamefont {Deng}}, \bibinfo {author} {\bibfnamefont {A.~A.}\ \bibnamefont {Lee}}, \bibinfo {author} {\bibfnamefont {A.}~\bibnamefont {Jain}},\ and\ \bibinfo {author} {\bibfnamefont {K.~A.}\ \bibnamefont {Persson}},\ }\bibfield  {title} {\bibinfo {title} {Matbench discovery -- a framework to evaluate machine learning crystal stability predictions},\ }\href {http://arxiv.org/abs/2308.14920} {\bibfield  {journal} {\bibinfo  {journal} {arXiv [cond-mat.mtrl-sci]}\ } (\bibinfo {year} {2023})},\ \Eprint {https://arxiv.org/abs/2308.14920} {arXiv:2308.14920 [cond-mat.mtrl-sci]} \BibitemShut {NoStop}%
\bibitem [{\citenamefont {Takamoto}\ \emph {et~al.}(2022{\natexlab{a}})\citenamefont {Takamoto}, \citenamefont {Shinagawa}, \citenamefont {Motoki}, \citenamefont {Nakago}, \citenamefont {Li}, \citenamefont {Kurata}, \citenamefont {Watanabe}, \citenamefont {Yayama}, \citenamefont {Iriguchi}, \citenamefont {Asano} \emph {et~al.}}]{takamoto2022towards}%
  \BibitemOpen
  \bibfield  {author} {\bibinfo {author} {\bibfnamefont {S.}~\bibnamefont {Takamoto}}, \bibinfo {author} {\bibfnamefont {C.}~\bibnamefont {Shinagawa}}, \bibinfo {author} {\bibfnamefont {D.}~\bibnamefont {Motoki}}, \bibinfo {author} {\bibfnamefont {K.}~\bibnamefont {Nakago}}, \bibinfo {author} {\bibfnamefont {W.}~\bibnamefont {Li}}, \bibinfo {author} {\bibfnamefont {I.}~\bibnamefont {Kurata}}, \bibinfo {author} {\bibfnamefont {T.}~\bibnamefont {Watanabe}}, \bibinfo {author} {\bibfnamefont {Y.}~\bibnamefont {Yayama}}, \bibinfo {author} {\bibfnamefont {H.}~\bibnamefont {Iriguchi}}, \bibinfo {author} {\bibfnamefont {Y.}~\bibnamefont {Asano}}, \emph {et~al.},\ }\bibfield  {title} {\bibinfo {title} {Towards universal neural network potential for material discovery applicable to arbitrary combination of 45 elements},\ }\href@noop {} {\bibfield  {journal} {\bibinfo  {journal} {Nature Communications}\ }\textbf {\bibinfo {volume} {13}},\ \bibinfo {pages} {2991} (\bibinfo {year} {2022}{\natexlab{a}})}\BibitemShut
  {NoStop}%
\bibitem [{\citenamefont {Takamoto}\ \emph {et~al.}(2023)\citenamefont {Takamoto}, \citenamefont {Okanohara}, \citenamefont {Li},\ and\ \citenamefont {Li}}]{takamoto2023towards}%
  \BibitemOpen
  \bibfield  {author} {\bibinfo {author} {\bibfnamefont {S.}~\bibnamefont {Takamoto}}, \bibinfo {author} {\bibfnamefont {D.}~\bibnamefont {Okanohara}}, \bibinfo {author} {\bibfnamefont {Q.-J.}\ \bibnamefont {Li}},\ and\ \bibinfo {author} {\bibfnamefont {J.}~\bibnamefont {Li}},\ }\bibfield  {title} {\bibinfo {title} {Towards universal neural network interatomic potential},\ }\href@noop {} {\bibfield  {journal} {\bibinfo  {journal} {Journal of Materiomics}\ }\textbf {\bibinfo {volume} {9}},\ \bibinfo {pages} {447} (\bibinfo {year} {2023})}\BibitemShut {NoStop}%
\bibitem [{\citenamefont {Miwa}\ \emph {et~al.}(2008)\citenamefont {Miwa}, \citenamefont {Martins},\ and\ \citenamefont {Fazzio}}]{miwa2008hydrogen}%
  \BibitemOpen
  \bibfield  {author} {\bibinfo {author} {\bibfnamefont {R.}~\bibnamefont {Miwa}}, \bibinfo {author} {\bibfnamefont {T.~B.}\ \bibnamefont {Martins}},\ and\ \bibinfo {author} {\bibfnamefont {A.}~\bibnamefont {Fazzio}},\ }\bibfield  {title} {\bibinfo {title} {Hydrogen adsorption on boron doped graphene: an ab initio study},\ }\href@noop {} {\bibfield  {journal} {\bibinfo  {journal} {Nanotechnology}\ }\textbf {\bibinfo {volume} {19}},\ \bibinfo {pages} {155708} (\bibinfo {year} {2008})}\BibitemShut {NoStop}%
\bibitem [{\citenamefont {Tabtimsai}\ \emph {et~al.}(2017)\citenamefont {Tabtimsai}, \citenamefont {Rakrai},\ and\ \citenamefont {Wanno}}]{tabtimsai2017hydrogen}%
  \BibitemOpen
  \bibfield  {author} {\bibinfo {author} {\bibfnamefont {C.}~\bibnamefont {Tabtimsai}}, \bibinfo {author} {\bibfnamefont {W.}~\bibnamefont {Rakrai}},\ and\ \bibinfo {author} {\bibfnamefont {B.}~\bibnamefont {Wanno}},\ }\bibfield  {title} {\bibinfo {title} {Hydrogen adsorption on graphene sheets doped with group 8b transition metal: a dft investigation},\ }\href@noop {} {\bibfield  {journal} {\bibinfo  {journal} {Vacuum}\ }\textbf {\bibinfo {volume} {139}},\ \bibinfo {pages} {101} (\bibinfo {year} {2017})}\BibitemShut {NoStop}%
\bibitem [{\citenamefont {Jain}\ and\ \citenamefont {Kandasubramanian}(2020)}]{jain2020functionalized}%
  \BibitemOpen
  \bibfield  {author} {\bibinfo {author} {\bibfnamefont {V.}~\bibnamefont {Jain}}\ and\ \bibinfo {author} {\bibfnamefont {B.}~\bibnamefont {Kandasubramanian}},\ }\bibfield  {title} {\bibinfo {title} {Functionalized graphene materials for hydrogen storage},\ }\href@noop {} {\bibfield  {journal} {\bibinfo  {journal} {Journal of Materials Science}\ }\textbf {\bibinfo {volume} {55}},\ \bibinfo {pages} {1865} (\bibinfo {year} {2020})}\BibitemShut {NoStop}%
\bibitem [{\citenamefont {Nakano}\ \emph {et~al.}(2006)\citenamefont {Nakano}, \citenamefont {Ohta}, \citenamefont {Yokoe}, \citenamefont {Doi},\ and\ \citenamefont {Tachibana}}]{NAKANO2006125}%
  \BibitemOpen
  \bibfield  {author} {\bibinfo {author} {\bibfnamefont {H.}~\bibnamefont {Nakano}}, \bibinfo {author} {\bibfnamefont {H.}~\bibnamefont {Ohta}}, \bibinfo {author} {\bibfnamefont {A.}~\bibnamefont {Yokoe}}, \bibinfo {author} {\bibfnamefont {K.}~\bibnamefont {Doi}},\ and\ \bibinfo {author} {\bibfnamefont {A.}~\bibnamefont {Tachibana}},\ }\bibfield  {title} {\bibinfo {title} {First-principle molecular-dynamics study of hydrogen adsorption on an aluminum-doped carbon nanotube},\ }\href {https://doi.org/https://doi.org/10.1016/j.jpowsour.2006.04.023} {\bibfield  {journal} {\bibinfo  {journal} {Journal of Power Sources}\ }\textbf {\bibinfo {volume} {163}},\ \bibinfo {pages} {125} (\bibinfo {year} {2006})},\ \bibinfo {note} {special issue including selected papers presented at the Second International Conference on Polymer Batteries and Fuel Cells together with regular papers}\BibitemShut {NoStop}%
\bibitem [{\citenamefont {Nagare}\ \emph {et~al.}(2012)\citenamefont {Nagare}, \citenamefont {Habale}, \citenamefont {Chacko},\ and\ \citenamefont {Ghosh}}]{nagare2012hydrogen}%
  \BibitemOpen
  \bibfield  {author} {\bibinfo {author} {\bibfnamefont {B.~J.}\ \bibnamefont {Nagare}}, \bibinfo {author} {\bibfnamefont {D.}~\bibnamefont {Habale}}, \bibinfo {author} {\bibfnamefont {S.}~\bibnamefont {Chacko}},\ and\ \bibinfo {author} {\bibfnamefont {S.}~\bibnamefont {Ghosh}},\ }\bibfield  {title} {\bibinfo {title} {Hydrogen adsorption on na--swcnt systems},\ }\href@noop {} {\bibfield  {journal} {\bibinfo  {journal} {Journal of Materials Chemistry}\ }\textbf {\bibinfo {volume} {22}},\ \bibinfo {pages} {22013} (\bibinfo {year} {2012})}\BibitemShut {NoStop}%
\bibitem [{\citenamefont {Verdinelli}\ \emph {et~al.}(2014)\citenamefont {Verdinelli}, \citenamefont {German}, \citenamefont {Luna}, \citenamefont {Marchetti}, \citenamefont {Volpe},\ and\ \citenamefont {Juan}}]{verdinelli2014theoretical}%
  \BibitemOpen
  \bibfield  {author} {\bibinfo {author} {\bibfnamefont {V.}~\bibnamefont {Verdinelli}}, \bibinfo {author} {\bibfnamefont {E.}~\bibnamefont {German}}, \bibinfo {author} {\bibfnamefont {C.~R.}\ \bibnamefont {Luna}}, \bibinfo {author} {\bibfnamefont {J.~M.}\ \bibnamefont {Marchetti}}, \bibinfo {author} {\bibfnamefont {M.~A.}\ \bibnamefont {Volpe}},\ and\ \bibinfo {author} {\bibfnamefont {A.}~\bibnamefont {Juan}},\ }\bibfield  {title} {\bibinfo {title} {Theoretical study of hydrogen adsorption on ru-decorated (8, 0) single-walled carbon nanotube},\ }\href@noop {} {\bibfield  {journal} {\bibinfo  {journal} {The Journal of Physical Chemistry C}\ }\textbf {\bibinfo {volume} {118}},\ \bibinfo {pages} {27672} (\bibinfo {year} {2014})}\BibitemShut {NoStop}%
\bibitem [{\citenamefont {Iijima}(1991)}]{iijima1991helical}%
  \BibitemOpen
  \bibfield  {author} {\bibinfo {author} {\bibfnamefont {S.}~\bibnamefont {Iijima}},\ }\bibfield  {title} {\bibinfo {title} {Helical microtubules of graphitic carbon},\ }\href@noop {} {\bibfield  {journal} {\bibinfo  {journal} {nature}\ }\textbf {\bibinfo {volume} {354}},\ \bibinfo {pages} {56} (\bibinfo {year} {1991})}\BibitemShut {NoStop}%
\bibitem [{\citenamefont {Frey}\ and\ \citenamefont {Doren}(2011)}]{tubegen34}%
  \BibitemOpen
  \bibfield  {author} {\bibinfo {author} {\bibfnamefont {J.~T.}\ \bibnamefont {Frey}}\ and\ \bibinfo {author} {\bibfnamefont {D.~J.}\ \bibnamefont {Doren}},\ }\href@noop {} {\bibinfo {title} {Tubegen 3.4 (web-interface)}},\ \bibinfo {howpublished} {\url{http://turin.nss.udel.edu/research/tubegenonline.html}} (\bibinfo {year} {2011}),\ \bibinfo {note} {university of Delaware, Newark DE}\BibitemShut {NoStop}%
\bibitem [{\citenamefont {Takamoto}\ \emph {et~al.}(2022{\natexlab{b}})\citenamefont {Takamoto}, \citenamefont {Izumi},\ and\ \citenamefont {Li}}]{takamoto2022teanet}%
  \BibitemOpen
  \bibfield  {author} {\bibinfo {author} {\bibfnamefont {S.}~\bibnamefont {Takamoto}}, \bibinfo {author} {\bibfnamefont {S.}~\bibnamefont {Izumi}},\ and\ \bibinfo {author} {\bibfnamefont {J.}~\bibnamefont {Li}},\ }\bibfield  {title} {\bibinfo {title} {Teanet: Universal neural network interatomic potential inspired by iterative electronic relaxations},\ }\href@noop {} {\bibfield  {journal} {\bibinfo  {journal} {Computational Materials Science}\ }\textbf {\bibinfo {volume} {207}},\ \bibinfo {pages} {111280} (\bibinfo {year} {2022}{\natexlab{b}})}\BibitemShut {NoStop}%
\bibitem [{\citenamefont {Kresse}\ and\ \citenamefont {Furthm{\"u}ller}(1996{\natexlab{a}})}]{kresse1996efficiency}%
  \BibitemOpen
  \bibfield  {author} {\bibinfo {author} {\bibfnamefont {G.}~\bibnamefont {Kresse}}\ and\ \bibinfo {author} {\bibfnamefont {J.}~\bibnamefont {Furthm{\"u}ller}},\ }\bibfield  {title} {\bibinfo {title} {Efficiency of ab-initio total energy calculations for metals and semiconductors using a plane-wave basis set},\ }\href@noop {} {\bibfield  {journal} {\bibinfo  {journal} {Computational materials science}\ }\textbf {\bibinfo {volume} {6}},\ \bibinfo {pages} {15} (\bibinfo {year} {1996}{\natexlab{a}})}\BibitemShut {NoStop}%
\bibitem [{\citenamefont {Kresse}\ and\ \citenamefont {Furthm{\"u}ller}(1996{\natexlab{b}})}]{kresse1996efficient}%
  \BibitemOpen
  \bibfield  {author} {\bibinfo {author} {\bibfnamefont {G.}~\bibnamefont {Kresse}}\ and\ \bibinfo {author} {\bibfnamefont {J.}~\bibnamefont {Furthm{\"u}ller}},\ }\bibfield  {title} {\bibinfo {title} {Efficient iterative schemes for ab initio total-energy calculations using a plane-wave basis set},\ }\href@noop {} {\bibfield  {journal} {\bibinfo  {journal} {Physical review B}\ }\textbf {\bibinfo {volume} {54}},\ \bibinfo {pages} {11169} (\bibinfo {year} {1996}{\natexlab{b}})}\BibitemShut {NoStop}%
\bibitem [{\citenamefont {Perdew}\ \emph {et~al.}(1996)\citenamefont {Perdew}, \citenamefont {Burke},\ and\ \citenamefont {Ernzerhof}}]{perdew1996generalized}%
  \BibitemOpen
  \bibfield  {author} {\bibinfo {author} {\bibfnamefont {J.~P.}\ \bibnamefont {Perdew}}, \bibinfo {author} {\bibfnamefont {K.}~\bibnamefont {Burke}},\ and\ \bibinfo {author} {\bibfnamefont {M.}~\bibnamefont {Ernzerhof}},\ }\bibfield  {title} {\bibinfo {title} {Generalized gradient approximation made simple},\ }\href@noop {} {\bibfield  {journal} {\bibinfo  {journal} {Physical review letters}\ }\textbf {\bibinfo {volume} {77}},\ \bibinfo {pages} {3865} (\bibinfo {year} {1996})}\BibitemShut {NoStop}%
\bibitem [{\citenamefont {Bl{\"o}chl}(1994)}]{blochl1994projector}%
  \BibitemOpen
  \bibfield  {author} {\bibinfo {author} {\bibfnamefont {P.~E.}\ \bibnamefont {Bl{\"o}chl}},\ }\bibfield  {title} {\bibinfo {title} {Projector augmented-wave method},\ }\href@noop {} {\bibfield  {journal} {\bibinfo  {journal} {Physical review B}\ }\textbf {\bibinfo {volume} {50}},\ \bibinfo {pages} {17953} (\bibinfo {year} {1994})}\BibitemShut {NoStop}%
\bibitem [{\citenamefont {Kresse}\ and\ \citenamefont {Joubert}(1999)}]{kresse1999ultrasoft}%
  \BibitemOpen
  \bibfield  {author} {\bibinfo {author} {\bibfnamefont {G.}~\bibnamefont {Kresse}}\ and\ \bibinfo {author} {\bibfnamefont {D.}~\bibnamefont {Joubert}},\ }\bibfield  {title} {\bibinfo {title} {From ultrasoft pseudopotentials to the projector augmented-wave method},\ }\href@noop {} {\bibfield  {journal} {\bibinfo  {journal} {Physical review b}\ }\textbf {\bibinfo {volume} {59}},\ \bibinfo {pages} {1758} (\bibinfo {year} {1999})}\BibitemShut {NoStop}%
\bibitem [{\citenamefont {Grimme}\ \emph {et~al.}(2010)\citenamefont {Grimme}, \citenamefont {Antony}, \citenamefont {Ehrlich},\ and\ \citenamefont {Krieg}}]{Grimme2010-yv}%
  \BibitemOpen
  \bibfield  {author} {\bibinfo {author} {\bibfnamefont {S.}~\bibnamefont {Grimme}}, \bibinfo {author} {\bibfnamefont {J.}~\bibnamefont {Antony}}, \bibinfo {author} {\bibfnamefont {S.}~\bibnamefont {Ehrlich}},\ and\ \bibinfo {author} {\bibfnamefont {H.}~\bibnamefont {Krieg}},\ }\bibfield  {title} {\bibinfo {title} {A consistent and accurate ab initio parametrization of density functional dispersion correction (dft-d) for the 94 elements h-pu},\ }\href {https://doi.org/10.1063/1.3382344} {\bibfield  {journal} {\bibinfo  {journal} {J. Chem. Phys.}\ }\textbf {\bibinfo {volume} {132}},\ \bibinfo {pages} {154104} (\bibinfo {year} {2010})}\BibitemShut {NoStop}%
\bibitem [{\citenamefont {Kostoglou}\ \emph {et~al.}(2021)\citenamefont {Kostoglou}, \citenamefont {Liao}, \citenamefont {Wang}, \citenamefont {Kondo}, \citenamefont {Tampaxis}, \citenamefont {Steriotis}, \citenamefont {Giannakopoulos}, \citenamefont {Kontos}, \citenamefont {Hinder}, \citenamefont {Baker}, \citenamefont {Bousser}, \citenamefont {Matthews}, \citenamefont {Rebholz},\ and\ \citenamefont {Mitterer}}]{KOSTOGLOU2021294}%
  \BibitemOpen
  \bibfield  {author} {\bibinfo {author} {\bibfnamefont {N.}~\bibnamefont {Kostoglou}}, \bibinfo {author} {\bibfnamefont {C.-W.}\ \bibnamefont {Liao}}, \bibinfo {author} {\bibfnamefont {C.-Y.}\ \bibnamefont {Wang}}, \bibinfo {author} {\bibfnamefont {J.~N.}\ \bibnamefont {Kondo}}, \bibinfo {author} {\bibfnamefont {C.}~\bibnamefont {Tampaxis}}, \bibinfo {author} {\bibfnamefont {T.}~\bibnamefont {Steriotis}}, \bibinfo {author} {\bibfnamefont {K.}~\bibnamefont {Giannakopoulos}}, \bibinfo {author} {\bibfnamefont {A.~G.}\ \bibnamefont {Kontos}}, \bibinfo {author} {\bibfnamefont {S.}~\bibnamefont {Hinder}}, \bibinfo {author} {\bibfnamefont {M.}~\bibnamefont {Baker}}, \bibinfo {author} {\bibfnamefont {E.}~\bibnamefont {Bousser}}, \bibinfo {author} {\bibfnamefont {A.}~\bibnamefont {Matthews}}, \bibinfo {author} {\bibfnamefont {C.}~\bibnamefont {Rebholz}},\ and\ \bibinfo {author} {\bibfnamefont {C.}~\bibnamefont {Mitterer}},\ }\bibfield  {title} {\bibinfo {title} {Effect of pt nanoparticle decoration on the h2 storage
  performance of plasma-derived nanoporous graphene},\ }\href {https://doi.org/https://doi.org/10.1016/j.carbon.2020.08.061} {\bibfield  {journal} {\bibinfo  {journal} {Carbon}\ }\textbf {\bibinfo {volume} {171}},\ \bibinfo {pages} {294} (\bibinfo {year} {2021})}\BibitemShut {NoStop}%
\bibitem [{\citenamefont {Berendsen}\ \emph {et~al.}(1984)\citenamefont {Berendsen}, \citenamefont {Postma}, \citenamefont {Van~Gunsteren}, \citenamefont {DiNola},\ and\ \citenamefont {Haak}}]{berendsen1984molecular}%
  \BibitemOpen
  \bibfield  {author} {\bibinfo {author} {\bibfnamefont {H.~J.}\ \bibnamefont {Berendsen}}, \bibinfo {author} {\bibfnamefont {J.~v.}\ \bibnamefont {Postma}}, \bibinfo {author} {\bibfnamefont {W.~F.}\ \bibnamefont {Van~Gunsteren}}, \bibinfo {author} {\bibfnamefont {A.}~\bibnamefont {DiNola}},\ and\ \bibinfo {author} {\bibfnamefont {J.~R.}\ \bibnamefont {Haak}},\ }\bibfield  {title} {\bibinfo {title} {Molecular dynamics with coupling to an external bath},\ }\href@noop {} {\bibfield  {journal} {\bibinfo  {journal} {The Journal of chemical physics}\ }\textbf {\bibinfo {volume} {81}},\ \bibinfo {pages} {3684} (\bibinfo {year} {1984})}\BibitemShut {NoStop}%
\bibitem [{\citenamefont {Larsen}\ \emph {et~al.}(2017)\citenamefont {Larsen}, \citenamefont {Mortensen}, \citenamefont {Blomqvist}, \citenamefont {Castelli}, \citenamefont {Christensen}, \citenamefont {Du\l{}ak}, \citenamefont {Friis}, \citenamefont {Groves}, \citenamefont {Hammer}, \citenamefont {Hargus}, \citenamefont {Hermes}, \citenamefont {Jennings}, \citenamefont {Jensen}, \citenamefont {Kermode}, \citenamefont {Kitchin}, \citenamefont {Kolsbjerg}, \citenamefont {Kubal}, \citenamefont {Kaasbjerg}, \citenamefont {Lysgaard}, \citenamefont {Maronsson}, \citenamefont {Maxson}, \citenamefont {Olsen}, \citenamefont {Pastewka}, \citenamefont {Peterson}, \citenamefont {Rostgaard}, \citenamefont {Schi\o{}tz}, \citenamefont {Sch{\"{u}}tt}, \citenamefont {Strange}, \citenamefont {Thygesen}, \citenamefont {Vegge}, \citenamefont {Vilhelmsen}, \citenamefont {Walter}, \citenamefont {Zeng},\ and\ \citenamefont {Jacobsen}}]{ASE}%
  \BibitemOpen
  \bibfield  {author} {\bibinfo {author} {\bibfnamefont {A.~H.}\ \bibnamefont {Larsen}}, \bibinfo {author} {\bibfnamefont {J.~J.}\ \bibnamefont {Mortensen}}, \bibinfo {author} {\bibfnamefont {J.}~\bibnamefont {Blomqvist}}, \bibinfo {author} {\bibfnamefont {I.~E.}\ \bibnamefont {Castelli}}, \bibinfo {author} {\bibfnamefont {R.}~\bibnamefont {Christensen}}, \bibinfo {author} {\bibfnamefont {M.}~\bibnamefont {Du\l{}ak}}, \bibinfo {author} {\bibfnamefont {J.}~\bibnamefont {Friis}}, \bibinfo {author} {\bibfnamefont {M.~N.}\ \bibnamefont {Groves}}, \bibinfo {author} {\bibfnamefont {B.}~\bibnamefont {Hammer}}, \bibinfo {author} {\bibfnamefont {C.}~\bibnamefont {Hargus}}, \bibinfo {author} {\bibfnamefont {E.~D.}\ \bibnamefont {Hermes}}, \bibinfo {author} {\bibfnamefont {P.~C.}\ \bibnamefont {Jennings}}, \bibinfo {author} {\bibfnamefont {P.~B.}\ \bibnamefont {Jensen}}, \bibinfo {author} {\bibfnamefont {J.}~\bibnamefont {Kermode}}, \bibinfo {author} {\bibfnamefont {J.~R.}\ \bibnamefont {Kitchin}}, \bibinfo {author}
  {\bibfnamefont {E.~L.}\ \bibnamefont {Kolsbjerg}}, \bibinfo {author} {\bibfnamefont {J.}~\bibnamefont {Kubal}}, \bibinfo {author} {\bibfnamefont {K.}~\bibnamefont {Kaasbjerg}}, \bibinfo {author} {\bibfnamefont {S.}~\bibnamefont {Lysgaard}}, \bibinfo {author} {\bibfnamefont {J.~B.}\ \bibnamefont {Maronsson}}, \bibinfo {author} {\bibfnamefont {T.}~\bibnamefont {Maxson}}, \bibinfo {author} {\bibfnamefont {T.}~\bibnamefont {Olsen}}, \bibinfo {author} {\bibfnamefont {L.}~\bibnamefont {Pastewka}}, \bibinfo {author} {\bibfnamefont {A.}~\bibnamefont {Peterson}}, \bibinfo {author} {\bibfnamefont {C.}~\bibnamefont {Rostgaard}}, \bibinfo {author} {\bibfnamefont {J.}~\bibnamefont {Schi\o{}tz}}, \bibinfo {author} {\bibfnamefont {O.}~\bibnamefont {Sch{\"{u}}tt}}, \bibinfo {author} {\bibfnamefont {M.}~\bibnamefont {Strange}}, \bibinfo {author} {\bibfnamefont {K.~S.}\ \bibnamefont {Thygesen}}, \bibinfo {author} {\bibfnamefont {T.}~\bibnamefont {Vegge}}, \bibinfo {author} {\bibfnamefont {L.}~\bibnamefont {Vilhelmsen}},
  \bibinfo {author} {\bibfnamefont {M.}~\bibnamefont {Walter}}, \bibinfo {author} {\bibfnamefont {Z.}~\bibnamefont {Zeng}},\ and\ \bibinfo {author} {\bibfnamefont {K.~W.}\ \bibnamefont {Jacobsen}},\ }\bibfield  {title} {\bibinfo {title} {The atomic simulation environment--a python library for working with atoms},\ }\href {https://doi.org/10.1088/1361-648X/aa680e} {\bibfield  {journal} {\bibinfo  {journal} {J. Phys. Condens. Matter}\ }\textbf {\bibinfo {volume} {29}},\ \bibinfo {pages} {273002} (\bibinfo {year} {2017})}\BibitemShut {NoStop}%
\bibitem [{\citenamefont {Myrzakhmetov}\ \emph {et~al.}(2025)\citenamefont {Myrzakhmetov}, \citenamefont {Shomenov}, \citenamefont {Sultanov}, \citenamefont {Wang},\ and\ \citenamefont {Mentbayeva}}]{MYRZAKHMETOV2025413}%
  \BibitemOpen
  \bibfield  {author} {\bibinfo {author} {\bibfnamefont {B.}~\bibnamefont {Myrzakhmetov}}, \bibinfo {author} {\bibfnamefont {T.}~\bibnamefont {Shomenov}}, \bibinfo {author} {\bibfnamefont {F.}~\bibnamefont {Sultanov}}, \bibinfo {author} {\bibfnamefont {Y.}~\bibnamefont {Wang}},\ and\ \bibinfo {author} {\bibfnamefont {A.}~\bibnamefont {Mentbayeva}},\ }\bibfield  {title} {\bibinfo {title} {Hydrogen adsorption on pristine and modified graphene: Dft insights into defects, doping, and decoration},\ }\href {https://doi.org/https://doi.org/10.1016/j.ijhydene.2025.04.083} {\bibfield  {journal} {\bibinfo  {journal} {International Journal of Hydrogen Energy}\ }\textbf {\bibinfo {volume} {126}},\ \bibinfo {pages} {413} (\bibinfo {year} {2025})}\BibitemShut {NoStop}%
\bibitem [{\citenamefont {Ghosh}\ and\ \citenamefont {Padmanabhan}(2017)}]{GHOSH201724237}%
  \BibitemOpen
  \bibfield  {author} {\bibinfo {author} {\bibfnamefont {S.}~\bibnamefont {Ghosh}}\ and\ \bibinfo {author} {\bibfnamefont {V.}~\bibnamefont {Padmanabhan}},\ }\bibfield  {title} {\bibinfo {title} {Beryllium-doped single-walled carbon nanotubes with stone-wales defects: A promising material to store hydrogen at room temperature},\ }\href {https://doi.org/https://doi.org/10.1016/j.ijhydene.2017.07.204} {\bibfield  {journal} {\bibinfo  {journal} {International Journal of Hydrogen Energy}\ }\textbf {\bibinfo {volume} {42}},\ \bibinfo {pages} {24237} (\bibinfo {year} {2017})}\BibitemShut {NoStop}%
\bibitem [{\citenamefont {Zhou}\ \emph {et~al.}(2009)\citenamefont {Zhou}, \citenamefont {Zu}, \citenamefont {Gao}, \citenamefont {Nie},\ and\ \citenamefont {Xiao}}]{zhou2009adsorption}%
  \BibitemOpen
  \bibfield  {author} {\bibinfo {author} {\bibfnamefont {Y.}~\bibnamefont {Zhou}}, \bibinfo {author} {\bibfnamefont {X.~T.}\ \bibnamefont {Zu}}, \bibinfo {author} {\bibfnamefont {F.}~\bibnamefont {Gao}}, \bibinfo {author} {\bibfnamefont {J.}~\bibnamefont {Nie}},\ and\ \bibinfo {author} {\bibfnamefont {H.}~\bibnamefont {Xiao}},\ }\bibfield  {title} {\bibinfo {title} {Adsorption of hydrogen on boron-doped graphene: A first-principles prediction},\ }\href@noop {} {\bibfield  {journal} {\bibinfo  {journal} {Journal of Applied Physics}\ }\textbf {\bibinfo {volume} {105}} (\bibinfo {year} {2009})}\BibitemShut {NoStop}%
\bibitem [{\citenamefont {Sawant}\ \emph {et~al.}(2021)\citenamefont {Sawant}, \citenamefont {Yadav}, \citenamefont {Banerjee}, \citenamefont {Patwardhan}, \citenamefont {Joshi},\ and\ \citenamefont {Dasgupta}}]{SAWANT202139297}%
  \BibitemOpen
  \bibfield  {author} {\bibinfo {author} {\bibfnamefont {S.~V.}\ \bibnamefont {Sawant}}, \bibinfo {author} {\bibfnamefont {M.~D.}\ \bibnamefont {Yadav}}, \bibinfo {author} {\bibfnamefont {S.}~\bibnamefont {Banerjee}}, \bibinfo {author} {\bibfnamefont {A.~W.}\ \bibnamefont {Patwardhan}}, \bibinfo {author} {\bibfnamefont {J.~B.}\ \bibnamefont {Joshi}},\ and\ \bibinfo {author} {\bibfnamefont {K.}~\bibnamefont {Dasgupta}},\ }\bibfield  {title} {\bibinfo {title} {Hydrogen storage in boron-doped carbon nanotubes: Effect of dopant concentration},\ }\href {https://doi.org/https://doi.org/10.1016/j.ijhydene.2021.09.183} {\bibfield  {journal} {\bibinfo  {journal} {International Journal of Hydrogen Energy}\ }\textbf {\bibinfo {volume} {46}},\ \bibinfo {pages} {39297} (\bibinfo {year} {2021})}\BibitemShut {NoStop}%
\bibitem [{\citenamefont {Lee}\ \emph {et~al.}(2009)\citenamefont {Lee}, \citenamefont {Ihm}, \citenamefont {Cohen},\ and\ \citenamefont {Louie}}]{lee2009calcium}%
  \BibitemOpen
  \bibfield  {author} {\bibinfo {author} {\bibfnamefont {H.}~\bibnamefont {Lee}}, \bibinfo {author} {\bibfnamefont {J.}~\bibnamefont {Ihm}}, \bibinfo {author} {\bibfnamefont {M.~L.}\ \bibnamefont {Cohen}},\ and\ \bibinfo {author} {\bibfnamefont {S.~G.}\ \bibnamefont {Louie}},\ }\bibfield  {title} {\bibinfo {title} {Calcium-decorated carbon nanotubes for high-capacity hydrogen storage: first-principles calculations},\ }\href@noop {} {\bibfield  {journal} {\bibinfo  {journal} {Physical Review B—Condensed Matter and Materials Physics}\ }\textbf {\bibinfo {volume} {80}},\ \bibinfo {pages} {115412} (\bibinfo {year} {2009})}\BibitemShut {NoStop}%
\bibitem [{\citenamefont {Lee}\ \emph {et~al.}(2010)\citenamefont {Lee}, \citenamefont {Ihm}, \citenamefont {Cohen},\ and\ \citenamefont {Louie}}]{lee2010calcium}%
  \BibitemOpen
  \bibfield  {author} {\bibinfo {author} {\bibfnamefont {H.}~\bibnamefont {Lee}}, \bibinfo {author} {\bibfnamefont {J.}~\bibnamefont {Ihm}}, \bibinfo {author} {\bibfnamefont {M.~L.}\ \bibnamefont {Cohen}},\ and\ \bibinfo {author} {\bibfnamefont {S.~G.}\ \bibnamefont {Louie}},\ }\bibfield  {title} {\bibinfo {title} {Calcium-decorated graphene-based nanostructures for hydrogen storage},\ }\href@noop {} {\bibfield  {journal} {\bibinfo  {journal} {Nano letters}\ }\textbf {\bibinfo {volume} {10}},\ \bibinfo {pages} {793} (\bibinfo {year} {2010})}\BibitemShut {NoStop}%
\bibitem [{\citenamefont {Pyle}\ \emph {et~al.}(2016)\citenamefont {Pyle}, \citenamefont {Gray},\ and\ \citenamefont {Webb}}]{pyle2016hydrogen}%
  \BibitemOpen
  \bibfield  {author} {\bibinfo {author} {\bibfnamefont {D.~S.}\ \bibnamefont {Pyle}}, \bibinfo {author} {\bibfnamefont {E.~M.}\ \bibnamefont {Gray}},\ and\ \bibinfo {author} {\bibfnamefont {C.}~\bibnamefont {Webb}},\ }\bibfield  {title} {\bibinfo {title} {Hydrogen storage in carbon nanostructures via spillover},\ }\href@noop {} {\bibfield  {journal} {\bibinfo  {journal} {International Journal of Hydrogen Energy}\ }\textbf {\bibinfo {volume} {41}},\ \bibinfo {pages} {19098} (\bibinfo {year} {2016})}\BibitemShut {NoStop}%
\bibitem [{\citenamefont {Kubas}(1988)}]{kubas1988molecular}%
  \BibitemOpen
  \bibfield  {author} {\bibinfo {author} {\bibfnamefont {G.~J.}\ \bibnamefont {Kubas}},\ }\bibfield  {title} {\bibinfo {title} {Molecular hydrogen complexes: coordination of a. sigma. bond to transition metals},\ }\href@noop {} {\bibfield  {journal} {\bibinfo  {journal} {Accounts of Chemical Research}\ }\textbf {\bibinfo {volume} {21}},\ \bibinfo {pages} {120} (\bibinfo {year} {1988})}\BibitemShut {NoStop}%
\bibitem [{\citenamefont {Brzhezinskaya}\ \emph {et~al.}(2014)\citenamefont {Brzhezinskaya}, \citenamefont {Shmatko}, \citenamefont {Yalovega}, \citenamefont {Krestinin}, \citenamefont {Bashkin},\ and\ \citenamefont {Bogoslavskaja}}]{brzhezinskaya2014electronic}%
  \BibitemOpen
  \bibfield  {author} {\bibinfo {author} {\bibfnamefont {M.}~\bibnamefont {Brzhezinskaya}}, \bibinfo {author} {\bibfnamefont {V.}~\bibnamefont {Shmatko}}, \bibinfo {author} {\bibfnamefont {G.}~\bibnamefont {Yalovega}}, \bibinfo {author} {\bibfnamefont {A.}~\bibnamefont {Krestinin}}, \bibinfo {author} {\bibfnamefont {I.}~\bibnamefont {Bashkin}},\ and\ \bibinfo {author} {\bibfnamefont {E.}~\bibnamefont {Bogoslavskaja}},\ }\bibfield  {title} {\bibinfo {title} {Electronic structure of hydrogenated carbon nanotubes studied by core level spectroscopy},\ }\href@noop {} {\bibfield  {journal} {\bibinfo  {journal} {Journal of Electron Spectroscopy and Related Phenomena}\ }\textbf {\bibinfo {volume} {196}},\ \bibinfo {pages} {99} (\bibinfo {year} {2014})}\BibitemShut {NoStop}%
\bibitem [{\citenamefont {Brzhezinskaya}\ \emph {et~al.}(2019)\citenamefont {Brzhezinskaya}, \citenamefont {Belenkov}, \citenamefont {Greshnyakov}, \citenamefont {Yalovega},\ and\ \citenamefont {Bashkin}}]{brzhezinskaya2019new}%
  \BibitemOpen
  \bibfield  {author} {\bibinfo {author} {\bibfnamefont {M.}~\bibnamefont {Brzhezinskaya}}, \bibinfo {author} {\bibfnamefont {E.}~\bibnamefont {Belenkov}}, \bibinfo {author} {\bibfnamefont {V.}~\bibnamefont {Greshnyakov}}, \bibinfo {author} {\bibfnamefont {G.}~\bibnamefont {Yalovega}},\ and\ \bibinfo {author} {\bibfnamefont {I.}~\bibnamefont {Bashkin}},\ }\bibfield  {title} {\bibinfo {title} {New aspects in the study of carbon-hydrogen interaction in hydrogenated carbon nanotubes for energy storage applications},\ }\href@noop {} {\bibfield  {journal} {\bibinfo  {journal} {Journal of Alloys and Compounds}\ }\textbf {\bibinfo {volume} {792}},\ \bibinfo {pages} {713} (\bibinfo {year} {2019})}\BibitemShut {NoStop}%
\bibitem [{\citenamefont {Talyzin}\ \emph {et~al.}(2011)\citenamefont {Talyzin}, \citenamefont {Luzan}, \citenamefont {Anoshkin}, \citenamefont {Nasibulin}, \citenamefont {Jiang}, \citenamefont {Kauppinen}, \citenamefont {Mikoushkin}, \citenamefont {Shnitov}, \citenamefont {Marchenko},\ and\ \citenamefont {Noreus}}]{talyzin2011hydrogenation}%
  \BibitemOpen
  \bibfield  {author} {\bibinfo {author} {\bibfnamefont {A.~V.}\ \bibnamefont {Talyzin}}, \bibinfo {author} {\bibfnamefont {S.}~\bibnamefont {Luzan}}, \bibinfo {author} {\bibfnamefont {I.~V.}\ \bibnamefont {Anoshkin}}, \bibinfo {author} {\bibfnamefont {A.~G.}\ \bibnamefont {Nasibulin}}, \bibinfo {author} {\bibfnamefont {H.}~\bibnamefont {Jiang}}, \bibinfo {author} {\bibfnamefont {E.~I.}\ \bibnamefont {Kauppinen}}, \bibinfo {author} {\bibfnamefont {V.~M.}\ \bibnamefont {Mikoushkin}}, \bibinfo {author} {\bibfnamefont {V.~V.}\ \bibnamefont {Shnitov}}, \bibinfo {author} {\bibfnamefont {D.~E.}\ \bibnamefont {Marchenko}},\ and\ \bibinfo {author} {\bibfnamefont {D.}~\bibnamefont {Noreus}},\ }\bibfield  {title} {\bibinfo {title} {Hydrogenation, purification, and unzipping of carbon nanotubes by reaction with molecular hydrogen: road to graphane nanoribbons},\ }\href@noop {} {\bibfield  {journal} {\bibinfo  {journal} {ACS nano}\ }\textbf {\bibinfo {volume} {5}},\ \bibinfo {pages} {5132} (\bibinfo {year}
  {2011})}\BibitemShut {NoStop}%
\bibitem [{\citenamefont {Lambie}\ \emph {et~al.}(2022)\citenamefont {Lambie}, \citenamefont {Steenbergen}, \citenamefont {Gaston},\ and\ \citenamefont {Paulus}}]{lambie2022clustering}%
  \BibitemOpen
  \bibfield  {author} {\bibinfo {author} {\bibfnamefont {S.}~\bibnamefont {Lambie}}, \bibinfo {author} {\bibfnamefont {K.~G.}\ \bibnamefont {Steenbergen}}, \bibinfo {author} {\bibfnamefont {N.}~\bibnamefont {Gaston}},\ and\ \bibinfo {author} {\bibfnamefont {B.}~\bibnamefont {Paulus}},\ }\bibfield  {title} {\bibinfo {title} {Clustering of metal dopants in defect sites of graphene-based materials},\ }\href@noop {} {\bibfield  {journal} {\bibinfo  {journal} {Physical Chemistry Chemical Physics}\ }\textbf {\bibinfo {volume} {24}},\ \bibinfo {pages} {98} (\bibinfo {year} {2022})}\BibitemShut {NoStop}%
\bibitem [{\citenamefont {Sun}\ \emph {et~al.}(2005)\citenamefont {Sun}, \citenamefont {Wang}, \citenamefont {Jena},\ and\ \citenamefont {Kawazoe}}]{sun2005clustering}%
  \BibitemOpen
  \bibfield  {author} {\bibinfo {author} {\bibfnamefont {Q.}~\bibnamefont {Sun}}, \bibinfo {author} {\bibfnamefont {Q.}~\bibnamefont {Wang}}, \bibinfo {author} {\bibfnamefont {P.}~\bibnamefont {Jena}},\ and\ \bibinfo {author} {\bibfnamefont {Y.}~\bibnamefont {Kawazoe}},\ }\bibfield  {title} {\bibinfo {title} {Clustering of ti on a c60 surface and its effect on hydrogen storage},\ }\href@noop {} {\bibfield  {journal} {\bibinfo  {journal} {Journal of the American Chemical Society}\ }\textbf {\bibinfo {volume} {127}},\ \bibinfo {pages} {14582} (\bibinfo {year} {2005})}\BibitemShut {NoStop}%
\end{thebibliography}%
\end{document}